\pdfoutput=1

\documentclass[11pt,dvipsnames]{article}

\usepackage[final]{coling}

\usepackage{times}
\usepackage{latexsym}

\usepackage[T1]{fontenc}

\usepackage[utf8]{inputenc}

\usepackage{microtype}

\usepackage{inconsolata}

\usepackage{float}
\usepackage{multicol}
\usepackage{multirow}
\usepackage{wrapfig}
\usepackage{graphicx}
\usepackage{transparent}
\usepackage{lipsum}
\usepackage{makecell}

\usepackage{amsmath,amsfonts,bm}

\def\eqref#1{equation~\ref{#1}}

\def\1{\bm{1}}

\DeclareMathAlphabet{\mathsfit}{\encodingdefault}{\sfdefault}{m}{sl}
\SetMathAlphabet{\mathsfit}{bold}{\encodingdefault}{\sfdefault}{bx}{n}

\usepackage{tcolorbox}
\newcommand\scalemath[2]{\scalebox{#1}{\mbox{\ensuremath{\displaystyle #2}}}}
\usepackage{url}            
\usepackage{booktabs}       
\usepackage{amsfonts}       
\usepackage{nicefrac}       
\usepackage{microtype}      
\usepackage{amsmath}
\usepackage{float}
\usepackage{caption}
\usepackage{graphicx,calc}
\usepackage{wrapfig}
\usepackage{color,soul}
\usepackage{enumitem}
\usepackage{multirow}
\usepackage{subcaption}
\usepackage{dcolumn}
\usepackage{tcolorbox}

\usepackage{colortbl}
\definecolor{aliceblue}{RGB}{178, 217, 245}

\definecolor{babyblue}{RGB}{217, 239, 251}

\definecolor{babypink}{RGB}{251, 231, 230}
\definecolor{mygreen}{HTML}{3cb44b}

\usepackage[linesnumbered,ruled,vlined]{algorithm2e}

\SetKwComment{Comment}{\color{green!50!black}\# }{}

\newcommand{\var}{\texttt}

\SetKwProg{Function}{def}{:}{}

\SetKwProg{For}{for}{:}{}
\SetKwProg{If}{if}{:}{}
\newcommand{\VarSty}[1]{\textnormal{\ttfamily\color{blue!90!black}#1}\unskip}

\usepackage{enumitem}

\usepackage[resetlabels,labeled]{multibib}
\newcites{Supp}{Supplementary Referrence}

\usepackage[capitalize]{cleveref}
\crefname{section}{Sec.}{Secs.}
\Crefname{section}{Section}{Sections}
\Crefname{table}{Table}{Tables}
\crefname{table}{Tab.}{Tabs.}
\title{InstructMol: Multi-Modal Integration for Building a Versatile and Reliable Molecular Assistant in Drug Discovery}

\author{
    He Cao\textsuperscript{1,2}\thanks{Work done during an internship at IDEA.},
    Zijing Liu\textsuperscript{1},
    Xingyu Lu\textsuperscript{1,3},
    {\bf Yuan Yao\textsuperscript{2}},
    {\bf Yu Li\textsuperscript{1}\thanks{\ \ Corresponding authors: Yu Li (\url{liyu@idea.edu.cn})}} \\
    \textsuperscript{1}International Digital Economy Academy (IDEA) \\
    \textsuperscript{2}Hong Kong University of Science and Technology\\
    \textsuperscript{3}Tsinghua Shenzhen International Graduate School, Tsinghua University \quad \\
    \texttt{hcaoaf@connect.ust.hk} \quad \texttt{luxy22@mails.tsinghua.edu.cn}\\
    \texttt{yuany@ust.hk} \quad \texttt{\{liuzijing,liyu\}@idea.edu.cn} \\
}

\begin{document}
\maketitle
\begin{abstract}
\vspace{-0.1in}

The rapid evolution of artificial intelligence in drug discovery encounters challenges with generalization and extensive training, yet Large Language Models (LLMs) offer promise in reshaping interactions with complex molecular data. Our novel contribution, \textbf{InstructMol}\footnote{\url{https://github.com/IDEA-XL/InstructMol}}, a multi-modal LLM, effectively aligns molecular structures with natural language via an instruction-tuning approach, utilizing a two-stage training strategy that adeptly combines limited domain-specific data with molecular and textual information. \textbf{InstructMol} showcases substantial performance improvements in drug discovery-related molecular tasks, surpassing leading LLMs and significantly reducing the gap with specialists, thereby establishing a robust foundation for a versatile and dependable drug discovery assistant.
\end{abstract}   
\section{Introduction}
\label{sec:intro}

\begin{figure*}
    \centering
    \includegraphics[width=\textwidth]{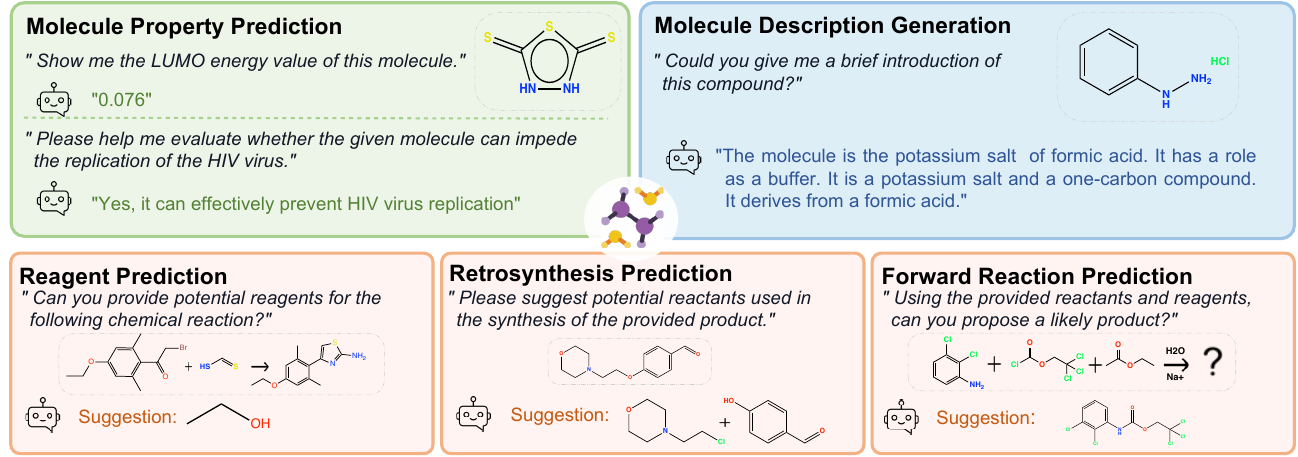}
    \captionof{figure}{Empowering LLMs with molecular modalities to unlock the drug discovery domain and serve as assistants in molecular research.}
    \label{fig:teaser}
    \vskip -0.1in
\end{figure*}

The drug discovery process, from target identification to clinical trials, requires substantial investments in time and expertise for optimized exploration of chemical spaces~\cite{chemicalspace}. Artificial intelligence-driven drug discovery (AIDD) facilitates a data-driven modeling approach~\cite{Kim2021ComprehensiveSO, Rifaioglu2018RecentAO, Askr2022DeepLI, feng2024bioactivity} and helps to understand the complex molecular space, reducing iterative testing and minimizing failure rates. Previous approaches involved employing task-specific models trained on labeled data, which had restricted adaptability and required laborious training for individual tasks. The advent of Large Language Models (LLMs~\cite{BERT, T5, GPT3}) like ChatGPT~\cite{Chatgpt}, trained through self-supervised learning on a large amount of unlabeled text data, has shown strong generalization capabilities across various tasks. Additionally, these models can attain professional-level proficiency in specific domains through proper fine-tuning. Hence, developing a ChatGPT-like molecular assistant AI can revolutionize human interactions with complex molecule structures. A unified model can address various needs, such as understanding molecule structures, answering drug-related queries, aiding synthesis planning, facilitating drug repurposing, etc., as shown in Figure~\ref{fig:teaser}.


Numerous studies have explored multimodal LLMs for visual understanding~\cite{llava, mPLUGOwlME, minigpt4}. However, when it comes to the domain of molecular research, there are several \textbf{challenges} that need to be addressed, including:
\begin{itemize}[noitemsep,topsep=0pt,leftmargin=*]
  \item Crafting a molecule representation integrates with LLMs alongside textual modalities;
  \item Requiring extensive datasets encompasses molecule structures, inherent properties, reactions, and annotations related to biological activities;
  \item Developing an effective training paradigm that guides LLMs in utilizing molecular representations and adapting to various tasks.
\end{itemize}

\noindent
Several prior studies~\cite{DrugChat, BioMedGPT, Mol-Instructions} have fine-tuned generalist LLMs to develop foundational models within the molecular domain. Despite their enhancement to the original generalist LLM, these preceding works have unveiled several \textbf{issues}: 
\begin{itemize}[noitemsep,topsep=0pt,leftmargin=*]
  \item  Insufficient alignment between modalities.
  \item  The consideration of an optimal molecular structure encoder remains unexplored.
  \item  A rudimentary design of the training pipeline neglects the update of LLMs' knowledge.
\end{itemize}
These issues lead to a significant disparity in the performance of current AI assistants across various practical tasks compared to traditional specialist models.

To address these problems, we introduce \textbf{InstructMol} (Figure~\ref{fig:overview}), a multi-modality instruction-tuning-based LLM. This model aligns molecular graphs and chemical sequential modalities with humans' natural language. Using a calibrated collection of molecule-related instruction datasets and a two-stage training scheme, \textbf{InstructMol} effectively leverages the pre-trained LLM and molecule graph encoder for molecule-text alignment. In the first alignment pretraining stage, we employ molecule-description pairs to train a lightweight and adaptable interface, which is designed to project the molecular node-level representation into the textual space that the LLM can understand. Subsequently, we finetune with multiple task-specific instructions. During this process, we freeze the molecule graph encoder and train low-rank adapters (LoRA~\cite{LoRA}) on the LLM to adapt our model to various scenarios. This efficient approach enables the seamless integration of molecular and textual information, promoting the development of versatile and robust cognitive abilities in the molecular domain.


To illustrate the capabilities of our model, we perform experiments that span three facets of drug discovery-related tasks, including compound property prediction, molecule description generation, and analysis of chemical reactions involving compounds. These tasks serve as robust benchmarks to assess the model's ability to deliver useful and accurate knowledge feedback in practical drug discovery scenarios. The results in all experiments consistently indicate that our model significantly improves the performance of LLMs in tasks related to the understanding and design of molecular compounds. 
Consequently, this advance effectively reduces the disparity with specialized models. Our main contributions can be summarized as follows:
\begin{itemize}[leftmargin=*,noitemsep,topsep=0pt]
    \item We introduce \textbf{InstructMol}, a molecular-related multi-modality LLM, representing a pioneering effort in bridging the gap between molecular and textual information.
    \vskip 0.5em
    \item In the context of a scarcity of high-quality annotated data in the drug discovery domain, our approach strives to efficiently extract molecular representations (targets on \textbf{Issue2}). Employing a two-stage instruction tuning paradigm enhances the LLM's understanding of molecular structural and sequential knowledge (targets on \textbf{Issue1} and \textbf{Issue3}).
    \vskip 0.5em
    \item InstructMol enables swift fine-tuning, generating lightweight checkpoints (used as plugins) for cross-modality tasks. It provides the flexibility to load or combine functionalities through plugins, retaining the open dialogue and reasoning capabilities of a general LLM. 
    \vskip 0.5em
    \item We evaluate our model through multiple practical assessments, demonstrating its substantial improvement compared to state-of-the-art LLMs. Our work lays the foundation for creating a versatile and reliable molecular research assistant in the drug discovery domain. 
\end{itemize}

\begin{table*}[htbp]
\centering
  \begin{minipage}[t]{0.66\linewidth}
  \setlength{\tabcolsep}{2.8 pt}
  \renewcommand\arraystretch{1.029}
    \centering
    \vskip -0.1in
    \includegraphics[width=1.0\linewidth]{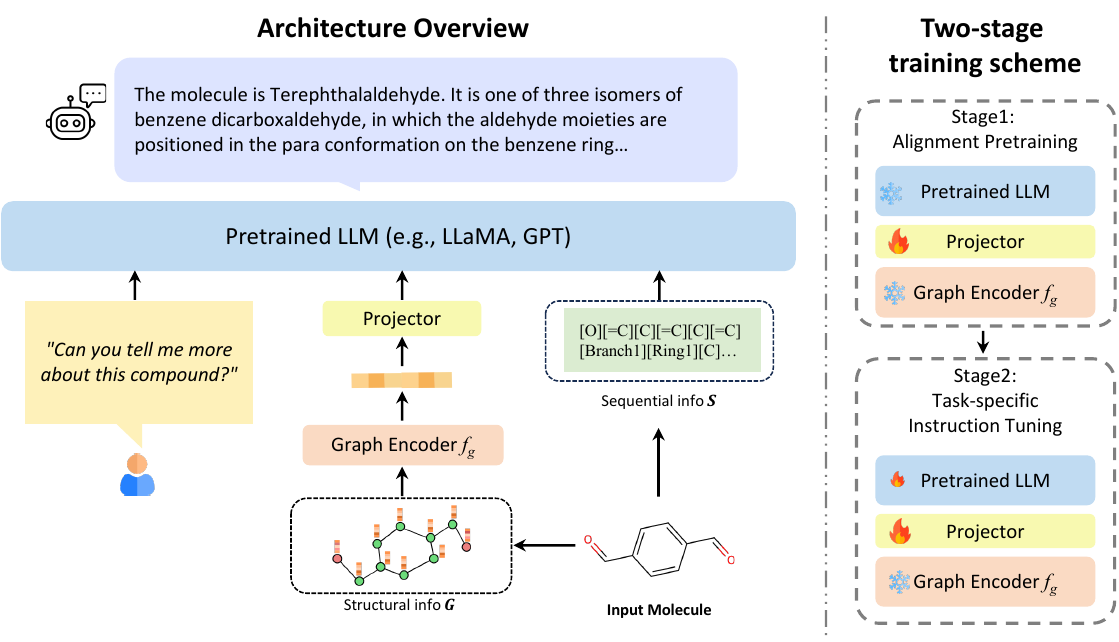}
    \vskip -0.1in
    \captionof{figure}{Overview of \textbf{InstructMol} model architecture design and two-stage training paradigm. The example molecule in the figure is \textit{Terephthalaldehyde}~\cite{Terephthalaldehyde} (CID 12173).}
    \label{fig:overview}
  \end{minipage}
  \hspace{6 pt}
  \begin{minipage}[t]{0.3\linewidth}
  \setlength{\tabcolsep}{2.8 pt}
  \renewcommand\arraystretch{1.029}
    \centering
    \vskip -0.1in
    \includegraphics[width=0.95\linewidth]{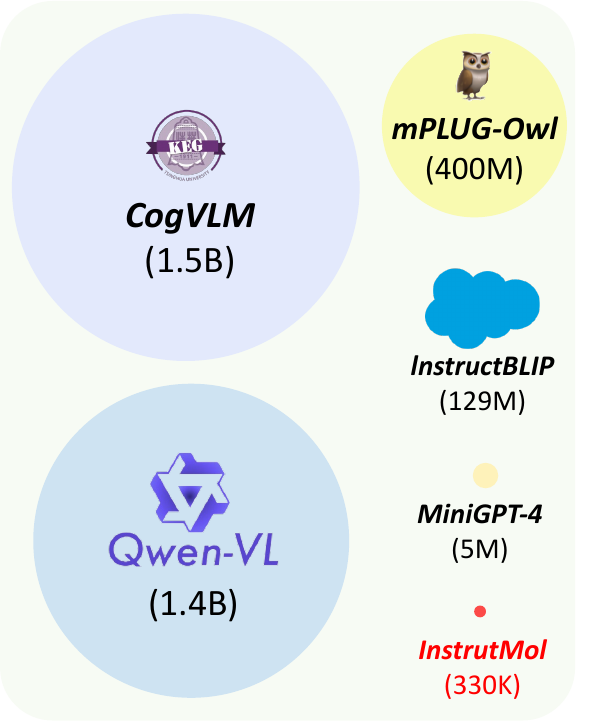}
    \captionof{figure}{\small Comparison of biomolecule-domain molecule-text dataset scale with existing general domain vision-language datasets.}
    \label{fig:dataset_scale}
  \end{minipage}
\end{table*}
\section{Related Work}
\label{sec:related_work}

\subsection{Multimodal Instruction Tuning}
There have been notable advancements in LLMs~\cite{Chatgpt, Llama, Llama2, vicuna, zeng2023glm-130b, palm2} achieved through scaling up model and data size. Consequently, LLMs have shown remarkable performances in zero/few-shot NLP tasks~\cite{Chatgpt, Wei2021FinetunedLM, InstructGPT}. A key technique in LLMs is instruction tuning, where pre-trained LLMs are fine-tuned on instruction-formatted datasets~\cite{Wei2021FinetunedLM}, allowing them to generalize to new tasks. Recently, with the emergence of large foundation models in various domains, several efforts have been made to transition from unimodal LLMs to multimodal LLMs (MLLMs)~\cite{gpt4v,llava,minigpt4,mPLUGOwlME,qwen-vl}. The primary research on multimodal instruction tuning (M-IT) includes the following~\cite{survey_mllm}: \textit{Constructing effective M-IT datasets} (adapting existing benchmarks datasets~\cite{minigpt4,llava,instructBLIP} or using self-instruction~\cite{llava,visionLLM,LLaVA-Med,pmc_vqa}), \textit{Bridging diverse modalities} (project-based~\cite{llava,LLaVA-Med,Pi2023DetGPTDW} and query-based~\cite{visionLLM,minigpt4,mPLUGOwlME}) and~\textit{Employing reliable evaluation methods} (GPT-scoring~\cite{llava, LLaVA-Med, X-LLM, LAVIN}, manual scoring~\cite{mPLUGOwlME,GPT4Tools}, or closed-set measurement~\cite{llava, LLaVA-Med, minigpt4, LAVIN, minigpt4, instructBLIP, X-LLM}). Most current MLLM research focuses on integrating vision and language while combining other modalities(e.g., graphs~\cite{GraphGPT, Liu2023TowardsGF}) with natural language remains nascent.

\subsection{Molecule Foundation Models}
The foundation models, trained on vast unlabeled data, serve as a paradigm for adaptable AI systems across diverse applications. 
In the single modality domain, researchers are exploring the molecule representations from diverse sources, such as 1D sequences (e.g., SMILES~\cite{ChemBERTa, Chemformer, SMILES-BERT}), 2D molecular graphs~\cite{Wang2021MolecularCL, Hu2019StrategiesFP, GraphCL}, 3D geometric conformations~\cite{3DInfomax, GraphMVP, 3DInfomax}, or textual information from biomedical literature~\cite{Gu2020DomainSpecificLM, BioBERT, SciBERT}. In the realm of multimodal analysis, research initiatives employ diverse approaches. These include encoder-decoder models to establish intermodal bridges~\cite{MolT5, Text+ChemT5, T5Chem}, joint generative modeling of SMILES and textual data~\cite{KV-PLM}, and the adoption of contrastive learning for integrating molecular knowledge across varying modalities~\cite{MoMu, MolFM, moleculeSTM, GIT-Mol}.

\subsection{Molecule-related LLMs}
Given the rapid progress in LLMs, some researchers are considering developing ChatGPT-like AI systems for drug discovery. Their goal is to offer guidance for optimizing lead compounds, accurately predicting drug interactions, and improving the comprehension of structure-activity relationships~\cite{DrugChat}. Several initiatives have already commenced to create instruction datasets within the biomolecular domain~\cite{Mol-Instructions, moleculeqa}. They aim to utilize instruction tuning techniques to enable LLMs, initially trained on general domain data, to acquire knowledge about biomolecular science~\cite{PMC-LLaMA, BioMedGPT}. Additionally, other researchers are investigating methods to align structural data with textual information, bridging the gap between biological data and natural language~\cite{BioMedGPT, DrugChat, presto}.

\vspace{0.1in}
\noindent
\textbf{Remark.} Our work involves molecule foundation models and multimodal language models (LLMs). It uses an efficient molecule graph encoder to capture structural information and integrates it with sequential data into a generalist LLM. \textbf{InstructMol} enables the LLM to understand molecule representations and generalize to various molecular tasks.
\section{Method}
\label{sec:method}

\subsection{Multimodal Instruction Tuning} 



Instruction tuning refers to finetuning pretrained LLMs on instruction datasets, enabling generalization to specific tasks by adhering to new instructions. Multimodal instruction tuning integrates modalities like images and graphs into an LLM, expanding the model's capability to accommodate multiple modalities. 

A multimodal instruction tuning sample comprises an instruction $I$ (e.g., \textit{"Describe the compound in detail"}) and an input-output pair. In the context of our study, the input is one or more modalities derived from a molecule (e.g., molecule graph and sequence), collectively denoted as $M$. The output $R$ represents the textual response to the instruction conditioned on the input. The model aims to predict an answer given the instruction and multimodal input: $\Tilde{R} = f(I,M;\theta)$, where $\theta$ are the parameters of MLLM. The training objective is typically the same auto-regressive objective as the LLM pre-training stage, which can be expressed as:
$
    \mathcal{L}(\theta) = -\sum_{i=1}^{L}\ \operatorname{log}\ p(R_i|I, M, R_{<i};\theta),
$
where $L$ is the target $R$'s token length.

\subsection{Construction of Molecular Instruction}

\noindent
\textbf{Data Collection.} In the field of biomolecular research, there is a noticeable scarcity of molecular datasets with comprehensive text annotations when compared to the vision-language domain, as depicted in Figure \ref{fig:dataset_scale}. 
While it is possible to construct instruction datasets in general domains by adapting benchmarks or using self-instruction, the application of these methods in the biomolecular domain presents challenges. This difficulty arises from two main factors: 1) biomolecular domain annotation demands expert knowledge and entails substantial complexity; 2) the knowledge within this domain spans a broad range of subjects, including structural biology, computational chemistry, and chemical synthesis processes.

In our efforts, we have gathered recent open-access text-molecule pairs datasets and also independently constructed a portion of instruction data suitable for property prediction. Table \ref{tab:data_collection} illustrates the composition of the data utilized during the two-stage training process.

\noindent
\textbf{Molecule Input.} We utilize both the structure and sequence information of a molecule. We encode the structural information of a molecule as a graph, denoted by $\mathcal{G}=(\mathcal{V}, \mathcal{E}, \mathcal{A}, \mathcal{X})$, where $\mathcal{V}$ is the set of atoms (nodes) and $|\mathcal{V}|=N$ is the total number of atoms. The set of edges $\mathcal{E}$ includes all chemical bonds, and $\mathcal{A}\in \mathbb{R}^{N\times N}$ is the adjacency matrix. Additionally, $\mathcal{X}\in \mathbb{R}^{N\times F}$ encompasses attributes associated with each node, where $F$ is the feature dimension. With a Graph Encoder $f_g$, we extract a graph representation $\mathbf{Z}_G \in \mathbb{R}^{N\times d}$ at the node level, effectively describing the inherent structure of the molecule. Simultaneously, we consider encoding the sequential information of the molecule, denoted as $S$, as a supplementary source of structural information. To enhance the robustness of sequential molecular descriptors and mitigate syntactic and semantic invalidity present in SMILES~\cite{SMILES}, we employ SELFIES~\cite{SELFIES} as $S$, which is designed for mapping each token to a distinct structure or reference.

\vskip 0.05in
\noindent
\textbf{Input Formulation.} We formulate a molecule-text pair ($\mathbf{X}_M$ \& $\mathbf{X}_c$) to the corresponding instruction-following version like \texttt{Human:} $\mathbf{X}_{I}$\texttt{<mol>}$\mathbf{X}_M$ \texttt{\color{black}<STOP>}  \texttt{Assistant:} $\mathbf{X}_A$\texttt{\color{black}<STOP>}. The $\mathbf{X}_M$ represents the molecule, including the molecule graph \colorbox{Apricot}{$\mathbf{X}_G$} and optionally the SELFIES \colorbox{SpringGreen}{$\mathbf{X}_S$}. $\mathbf{X}_I$ denotes for the instruction and $\mathbf{X}_A$ is the answer. For a given answer sequence of length $L$, our optimization objective is to maximize the probability of generating the target answers $X_A$ by maximizing:
\begin{equation}
\scalemath{0.88}{
p(\mathbf{X}_A|\mathbf{X}_M, \mathbf{X}_{I})=\prod_{i=1}^{L}p_{\theta}(x_i|\colorbox{Apricot}{$\mathbf{X}_G$}\mathbin\Vert \colorbox{SpringGreen}{$\mathbf{X}_S$}, \mathbf{X}_{I}, \mathbf{X}_{A,<i})}.
\end{equation}
To diversify $\mathbf{X}_{I}$, we craft clear task descriptions and use GPT-3.5-turbo to generate varied questions, enhancing instructions' robustness. Note that we simply concatenate \colorbox{Apricot}{$\mathbf{X}_G$} and \colorbox{SpringGreen}{$\mathbf{X}_S$} along the length-dimension. More complex fusion methods require additional loss designs for supervision~\cite{GIT-Mol, MolFM}, but here we prioritize simplicity.

\subsection{Architecture}
\textbf{Molecular Encoder.}
The molecular graph encoder, $f_g$, needs to efficiently extract node representations while preserving the molecular graph's connectivity information. It is crucial that $f_g$ inherently establishes a pre-alignment in the representation space with the text space to facilitate $\mathbf{Z}_G$ in the following alignment stage. Taking inspiration from common practices in the Vision Large Language Models (VLLM) domain~\cite{qwen-vl, llava, mPLUGOwlME}, where models like ViT initialized from CLIP~\cite{CLIP} serve as vision encoders, we optimize for MoleculeSTM's graph encoder as $f_g$~\cite{moleculeSTM}, instead of GraphMVP used by prior methodologies~\cite{DrugChat, BioMedGPT}. The MoleculeSTM graph encoder model is obtained through molecular-textual contrastive training, mitigating the requirement for an extensive amount of paired data during training to align different modalities.

\vskip 0.05in
\noindent
\textbf{Light-weight Alignment Projector.} To map graph features into the word embedding space, we utilize a trainable projection matrix $\mathbf{W}$ to transform $\mathbf{Z}_G$ into $\mathbf{X}_G$, ensuring that it has the same dimension as the word embedding space. 
Since the selected $f_g$ has undergone partial alignment with the text through contrastive training, we believe a straightforward linear projection will meet the subsequent alignment needs.
For approaches like gated cross-attention~\cite{Flamingo}, Q-former~\cite{BLIP2}, or position-aware vision-language adapters~\cite{qwen-vl}, they require a large number of pairs for pretraining alignment, which is typically unavailable in the biomolecular domain.
We therefore do not explore these more complex alignment methods.

\vskip 0.05in
\noindent
\textbf{Large Language Model.} InstructMol incorporates a pre-trained LLM as its foundational component. We optimize for Vicuna-7B~\cite{vicuna} as the initialized weights, which is derived from LLaMA~\cite{Llama} through supervised instruction finetuning.

\subsection{Two-Stage of Instruction Tuning}
As illustrated in Figure~\ref{fig:overview}, the training process of InstructMol consists of two stages: alignment pretraining and instruction fine-tuning training.

\noindent
\textbf{Alignment Pretraining.}
In the first stage, we aim to align the modality of molecules with text, ensuring that the LLMs can perceive both the structural and sequential information of molecules and integrate molecular knowledge into their internal capabilities. 

We primarily employ a dataset consisting of molecule-text pairs sourced from PubChem~\cite{PubChem}. Each molecule structure is associated with a textual description elucidating chemical and physical properties or high-level bioactivity information. The construction of the PubChemDataset predominantly follows the MoleculeSTM~\cite{moleculeSTM} pipeline. We meticulously remove molecules with invalid descriptions and syntactic errors in their molecular descriptors. To ensure fairness, we also eliminate compounds that might appear in the downstream molecule-caption test set. This results in a dataset of 330K molecule-text pairs. Subsequently, we adopt a self-instruction-like approach to generate a diverse set of task descriptions as instructions. 

During training, to prevent overfitting and leverage pre-trained knowledge, we freeze both the graph encoder and LLM, focusing solely on fine-tuning the alignment projector. After a few epochs of training, our aim is that the projector has successfully learned to map graph representations to graph tokens, aligning effectively with text tokens.

\vskip 0.05in
\noindent
\textbf{Task-specific Instruction Tuning.} In the second stage, we target three distinct downstream scenarios. We advocate for task-specific instruction tuning to address the particular constraints inherent in various drug-discovery-related tasks. For \textit{compound property prediction}, we utilize the quantum mechanics properties instruction dataset from \citet{Mol-Instructions} for regression prediction and the MoleculeNet dataset~\cite{MoleculeNet} for property classification. For \textit{chemical reaction analysis}, we incorporate forward reaction prediction, retrosynthesis analysis, and reagent prediction tasks, all derived from \citet{Mol-Instructions}. To assess the model's proficiency in translating between natural language and molecular expression, we integrate ChEBI-20~\cite{Text2Mol} for the \textit{molecule description generation task}. For each task, corresponding instruction templates are designed.

During training, we utilize the checkpoint of the alignment projector that was trained in the first stage as initialization. We only keep the molecular encoder $f_g$ frozen and continue to update the pre-trained weights of the projector and the LLM. To adapt the LLM effectively for diverse tasks, we employ low-rank adaptation (i.e., LoRA~\cite{LoRA}), opting against full-tuning to mitigate potential forgetting issues. In practical applications, we have the flexibility to substitute different adaptors based on specific scenario requirements or combine multiple adaptors to integrate knowledge, thereby showcasing the model's modularization capabilities. Moreover, LoRA allows the LLM to retain the inherent capacity for common-sense reasoning in dialogue (as shown in Table~\ref{tab:model_performance_combined}).

\section{Experiments}
We use a graph neural network as the molecule graph encoder ($f_g$) which is initialized with the MoleculeSTM graph encoder, pre-trained through molecular graph-text contrastive learning. We employ Vicuna-v-1.3-7B~\cite{vicuna} as the base LLM. More specifically, \textbf{InstructMol+GS} denotes we inject both molecular graph tokens and sequence tokens into the input, while \textbf{InstructMol+G} means only incorporates graph tokens. For Instruct-S, which utilizes only a 1D molecular sequence as input, it corresponds to the fine-tuning of the base large language model, Vicuna-7B, directly on downstream tasks. In the following sections, the results of Vicuna-v1.3-7B will consistently be used to represent the performance of Instruct-S. Implementation details about model settings and training hyper-parameters can be referred to Appendix \ref{sec:implementation_details}.

\subsection{Property Prediction Task} 
\textbf{Experiment Setup.} Property prediction intends to forecast a molecule's intrinsic physical and chemical properties from its structural or sequential characteristics. In the context of the regression task, we undertake experiments on the Property Prediction dataset from \citet{Mol-Instructions}, where the objective is to predict the quantum mechanic's properties of a given molecule, specifically including HOMO, LUMO, and the HOMO-LUMO gap~\cite{ramakrishnan2014quantum}. For the classification task, we incorporate three binary classification datasets of molecular biological activity, namely BACE, BBBP, and HIV. In classification, all dataset samples are converted into an instruction format and we use the recommended splits from~\cite{deepchem}. Each item comprises an instruction explaining the property for prediction and the representation of the molecule. Subsequently, models are tasked with generating a single prediction (\textit{``yes"} or \textit{``no"}). Scaffold splits are used for the classification task, and the experiments are conducted with three random seeds, yielding low variances in the reported mean values.

\begin{table}[htbp]
\centering
    \centering
    \scriptsize
    \scalebox{0.92}{
    \begin{tabular}{lccccc}
        \toprule
        \textsc{Method} &\textsc{HOMO} $\downarrow$ &\textsc{LUMO} $\downarrow$  &$\Delta{\epsilon}$ $\downarrow$ &\textsc{Avg} $\downarrow$\\
        \midrule 
        \rowcolor[RGB]{234, 238, 234} \multicolumn{5}{l}{\textit{LLM Based Generalist Models}} \\
        \rowcolor[RGB]{255, 255, 255} Alpaca$^\dagger$~\cite{alpaca} & - & - & - & 322.109\\
        Baize$^\dagger$~\cite{baize} & - & - & - & 261.343\\
        Galactica$^\dagger$~\cite{Galactica} & - & - & - & 0.568\\
        LLama-2-7B (5-shot ICL) & 0.7367 & 0.8641 & 0.5152 & 0.7510 \\
        Vicuna-13B (5-shot ICL) & 0.7135 & 3.6807 & 1.5407 & 1.9783 \\
        Mol-Instruction & 0.0210 & 0.0210 & 0.0203 & 0.0210 \\
        \textbf{InstructMol-G} &  0.0060 & 0.0070 & 0.0082 & 0.0070 \\
        \textbf{InstructMol-GS} & \textbf{0.0048} &\textbf{0.0050}	&\textbf{0.0061} & \textbf{0.0050}\\
        \bottomrule
    \end{tabular}
    }
    \caption{Results (MAE in hartree unit) for QM9 property regression tasks. $\dagger$ denotes few-shot in-context learning (ICL) results from~\citet{Mol-Instructions}. $\Delta{\epsilon}$ denotes HOMO-LUMO energy gap.}
    \label{tab:property_pred_regression}
  \end{table}

\vskip -0.2in
\begin{table}[!htp]
    \centering
    \scriptsize
    \scalebox{0.9}{
    \begin{tabular}{lccc}
        \toprule
        \textsc{Method} &BACE $\uparrow$  &BBBP $\uparrow$ &HIV $\uparrow$ \\
        \textsc{\# Molecules} &1513 &2039 &41127 \\
        \midrule
        \rowcolor[RGB]{234, 238, 234} \multicolumn{4}{l}{\textit{Specialist Models}} \\
        ChemBERTa v2~\cite{chemberta2} &73.5 &69.8 &79.3 \\
        DMP(TF+GNN)~\cite{DMP}	&\underline{89.4}	&\underline{77.8}	&\underline{81.4} \\
        KV-PLM~\cite{KV-PLM} &78.5 &70.5 &71.8 \\
        GraphCL~\cite{GraphCL} &75.3 &69.7 &78.5 \\
        GraphMVP-C~\cite{GraphMVP} &81.2 &72.4 &77.0 \\
        MoMu~\cite{MoMu} &76.7 &70.5 & 75.9 \\
        MolFM~\cite{MolFM} &83.9 &{72.9} &78.8 \\
        Uni-Mol~\cite{UniMol} &{85.7} &{72.9 }&{80.8} \\
        \midrule 
        \rowcolor[RGB]{234, 238, 234} \multicolumn{4}{l}{\textit{LLM Based Generalist Models}} \\
        \rowcolor[RGB]{255, 255, 255} Galactica-6.7B & 58.4 & 53.5 & 72.2 \\
        Vicuna-v1.5-13b-16k (4-shot) & 49.2 & 52.7 & 50.5 \\
        Vicuna-v1.3-7B$^*$  & 68.3 & 60.1 & 58.1 \\
        LLama-2-7B-chat$^*$ & 74.8 & 65.6 & 62.3 \\
        MolCA(1D)~\cite{MolCA} & 	79.3	&70.8	&- \\
        MolCA(1D + 2D)~\cite{MolCA} & 	79.8	&70.0	&- \\
        \textbf{Instruct-G} & \textbf{84.3} ($\pm$0.6) & 68.6 ($\pm$0.3) & \textbf{74.0} ($\pm$0.1) \\
        \textbf{Instruct-GS} & 82.1 ($\pm$0.1) & \textbf{72.4} ($\pm$0.3) & 68.9 ($\pm$0.3) \\
        \bottomrule
    \end{tabular}
    }
    \caption{
    ROC-AUC of molecular property prediction tasks (classification) on MoleculeNet~\cite{MoleculeNet} benchmarks. $^*$: use LoRA tuning. We indicate the best performance among domain specialist models by underlining the results, while the best performance among LLM-based generalist models is highlighted in bold.}
    \label{tab:property_pred_classification}
\end{table}

\begin{table*}[!htp]
\centering
\small
\setlength{\tabcolsep}{2mm}{
\scalebox{0.85}{
\begin{tabular}{lcccccc}
\toprule
\textsc{Model}
&\textsc{BLEU-2}$\uparrow$  & \textsc{BLEU-4}$\uparrow$  & \textsc{ROUGE-1}$\uparrow$  & \textsc{ROUGE-2}$\uparrow$  & \textsc{ROUGE-L}$\uparrow$ & \textsc{METEOR}$\uparrow$ \\
\midrule[1.1pt]
\rowcolor[RGB]{234, 238, 234} \multicolumn{7}{l}{\textit{Specialist Models}} \\
MolT5-base~\cite{MolT5}        & 0.540 &0.457 &0.634 &0.485 &0.568 &0.569 \\
MoMu (MolT5-base)~\cite{MoMu} & 0.549 &0.462 &  -    &  -    &  -    &0.576 \\
MolFM (MolT5-base)~\cite{MolFM}&0.585 &0.498 &0.653 &0.508 &0.594 &0.607 \\
MolXPT~\cite{molxpt}          &0.594  &0.505 &0.660 &0.511 &0.597 &0.626 \\
GIT-Mol-graph~\cite{GIT-Mol}  & 0.290 &0.210 &0.540 &0.445 &0.512 & 0.491 \\
GIT-Mol-SMILES~\cite{GIT-Mol} & 0.264 &0.176 &0.477 &0.374 &0.451 &0.430  \\
GIT-Mol-(graph+SMILES)~\cite{GIT-Mol} & 0.352 & 0.263 &0.575 &0.485 &0.560 &0.430 \\
$\text{MolCA, Galac}_{\text{1.3B}}$~\cite{MolCA} & 0.620 & 0.531 & 0.681 & 0.537 & 0.618 & \underline{0.651} \\
Text+Chem T5-augm-base~\cite{Text+ChemT5} &\underline{0.625} &\underline{0.542} &\underline{0.682} &\underline{0.543} &\underline{0.622} & 0.648 \\
\rowcolor[RGB]{234, 238, 234} \multicolumn{7}{l}{\textit{Retrieval Based LLMs}} \\
GPT-3.5-turbo (10-shot MolReGPT)~\cite{MolReGPT} &0.565 &0.482 &0.623 &0.450 &0.543 &0.585 \\
GPT-4-0314 (10-shot MolReGPT)~\cite{MolReGPT} &0.607 &0.525 &0.634 &0.476 &0.562 &0.610 \\
\midrule
\rowcolor[RGB]{234, 238, 234} \multicolumn{7}{l}{\textit{LLM Based Generalist Models}} \\
GPT-3.5-turbo (zero-shot)~\cite{MolReGPT} & 0.103 &0.050 &0.261 &0.088 &0.204 &0.161 \\
BioMedGPT-10B~\cite{BioMedGPT}          &0.234 &0.141  &0.386  &0.206  &0.332  &0.308  \\
Mol-Instruction~\cite{Mol-Instructions} &0.249 &0.171  &0.331  &0.203  &0.289  &0.271  \\
\textbf{InstructMol-G}         &0.466	&0.365	&0.547	&0.365	&0.479	&0.491	\\
\textbf{InstructMol-GS} &\textbf{0.475} &\textbf{0.371} &\textbf{0.566} &\textbf{0.394} &\textbf{0.502} &\textbf{0.509}\\
\bottomrule
\end{tabular}
}}
\caption{
Results of molecular description generation task on the test split of ChEBI-20.
}
\label{tab:molcap}
\end{table*}

\noindent
\textbf{Results.} Our models are compared against baselines on the test set for regression, measured by Mean Absolute Error (MAE) in Table \ref{tab:property_pred_regression}. Compared to previous single-modal instruction-tuned LLM-based methods~\cite{Mol-Instructions}, InstructMol demonstrates a further improvement in the regression task. ROC-AUC scores for classification outcomes are presented in Table \ref{tab:property_pred_classification}. In comparison to LLM-based generalist models, both the Galactica~\cite{Galactica} series models trained on an extensive scientific literature dataset and the single-modality LLM fine-tuned with task-specific instructions~\cite{Mol-Instructions}, InstructMol demonstrates consistent improvements in accuracy across the three task datasets. However, our predictive results still exhibit some disparity compared to expert models~\cite{UniMol, GraphMVP} specifically trained on a vast molecule structure dataset. Further, InstructMol performs worse than GIN on the imbalanced HIV dataset with a long-tail distribution. Previous research~\cite{Kandpal2022LargeLM} highlights LLMs' challenges in learning long-tail knowledge. To tackle this, strategies like resampling or class reweighting can be employed.

\subsection{Molecule Description Generation Task} 

\textbf{Experiment Setup.} 
Molecule description generation encapsulates a comprehensive molecule depiction, covering its structure, properties, biological activity, and applications based on molecular descriptors. This task is more complex than classification or regression, providing a robust measure of the model's understanding of molecules. 
We convert the training subset of the ChEBI-20 dataset~\cite{Text2Mol} into an instructional format and subsequently perform fine-tuning based on these instructions. Our assessment uses evaluation metrics aligned with \cite{MolT5}. 

\noindent
\textbf{Baselines.} Three kinds of models are used as baselines, including: 1) MolT5-like expert models~\cite{MolT5, molxpt} and the models employing MolT5 as a decoder~\cite{MoMu, MolFM, GIT-Mol, Text+ChemT5}, 2) models based on retrieval methods that utilize ChatGPT/GPT-4 as a foundational component~\cite{MolReGPT}, 3) other models derived through instruction-tuning with LLMs to achieve generalist unimodal~\cite{Mol-Instructions} and multi-modalities~\cite{BioMedGPT} capabilities.

\noindent
\textbf{Results.} Table \ref{tab:molcap} presents the overall results for molecule description generation. Our model outperforms other generalist LLM-based models in generating precise, contextually relevant molecule descriptions. We observe that incorporating both molecule structural information and sequential information in the input yields higher-quality results ($\sim$2\% improvement) than providing structural information alone. While expert models demonstrate better efficacy in comparison, it is noteworthy that they are constrained by their training schemes and lack the versatile capabilities inherent in our approach. 
Retrieval methods, supported by ChatGPT/GPT-4, demonstrate strong capabilities. Our future efforts will focus on integrating these methods to improve the accuracy and credibility of generated content.

\subsection{Chemical Reaction-related task} 
\textbf{Experiment Setup.} Traditionally, identifying chemical reactions relied on intuition and expertise. Integrating deep learning for predicting reactions can accelerate research and improve drug discovery. The general format of a chemical reaction is \texttt{"reactant $\rightarrow$ reagent $\rightarrow$ product"}. Here we mainly focus on three tasks: 1) \textit{Forward Reaction Prediction}: predict the probable product(s) given specific reactants and reagents; 2) \textit{Reagent Prediction}: ascertain the suitable catalysts, solvents, or ancillary substances required for a specific chemical reaction given reactant(s) and product(s); 3) \textit{Retrosynthesis}: anticipate deducing potential precursor molecule(s) from given product(s). 

We utilize the dataset sourced from~\citet{Mol-Instructions}, training it on the pre-defined training split and evaluating its performance on the test set. The performance is assessed by metrics like Fingerprint Tanimoto Similarity (FTS), BLEU, Exact Match, and Levenshtein distance to measure the similarity between ground truth and prediction. We also measure the validity of predicted molecules using RDKit.

\noindent
\textbf{Results.} Table \ref{tab:chemical_reaction} reports the outcomes of tasks related to chemical reactions. It is evident that \textbf{InstructMol} outperforms the baselines significantly. The results obtained by generalist LLMs are derived from \citet{Mol-Instructions}, and they exhibit a pronounced inability to comprehend any chemical reaction prediction task, struggling to generate valid molecule(s) as answers. Mol-Instruction~\cite{Mol-Instructions}, employing Llama2~\cite{Llama2} as the base LLM, is jointly trained on multiple molecule-oriented instruction datasets. In addition, we supplement this by adopting the same training settings but exclusively training on chemical reaction-related datasets. Through comparison, InstructMol, as a multi-modality LLM, demonstrates a superior understanding of the task compared to single-modality models, confirming its effectiveness as a chemical reaction assistant.

\begin{table*}[!htp]
\centering
\small
\setlength{\tabcolsep}{2mm}{
\scalebox{0.78}{
\begin{tabular}{lccccccc}
\toprule
\textsc{Model}
&\textsc{Exact}$\uparrow$  & \textsc{BLEU}$\uparrow$  & \textsc{Levenshtein}$\downarrow$  & \textsc{RDK FTS}$\uparrow$  & \textsc{MACCS FTS}$\uparrow$ & \textsc{Morgan FTS}$\uparrow$ & \textsc{Validity}$\uparrow$ \\

\midrule[1.1pt]
\rowcolor[RGB]{234, 238, 234}
\multicolumn{8}{l}{\textit{Reagent Prediction}} \\
Alpaca$^\dagger$~\cite{alpaca} & 0.000 & 0.026 & 29.037 & 0.029 & 0.016 & 0.001 & 0.186 \\
Baize$^\dagger$~\cite{baize} & 0.000 & 0.051 & 30.628 & 0.022 & 0.018 & 0.004 & 0.099 \\
ChatGLM$^\dagger$~\cite{zeng2023glm-130b} & 0.000 & 0.019 & 29.169 & 0.017 & 0.006 & 0.002 & 0.074 \\
LLama$^\dagger$~\cite{Llama} & 0.000 & 0.003 & 28.040 & 0.037 & 0.001 & 0.001 & 0.001 \\
Vicuna$^\dagger$~\cite{vicuna} & 0.000 & 0.010 & 27.948 & 0.038 & 0.002 & 0.001 & 0.007 \\
Mol-Instruction~\cite{Mol-Instructions} & 0.044 & 0.224 & 23.167 & 0.237 & 0.364 & 0.213 & 1.000 \\
LLama-7b$^*$~\cite{Llama}(LoRA) &0.000	&0.283	&53.510	&0.136	&0.294	&0.106	&1.000 \\
\midrule
\textbf{InstructMol-G} & 0.070 & \textbf{0.890} &24.732	&\textbf{0.469}	&\textbf{0.691}	&\textbf{0.426}	&1.000 \\
\textbf{InstructMol-GS} & \textbf{0.129} &0.610	&\textbf{19.664}	&0.444	&0.539	&0.400	&1.000 \\

\midrule[1.1pt]
\rowcolor[RGB]{234, 238, 234}
\multicolumn{8}{l}{\textit{Forward Reaction Prediction}} \\
Alpaca$^\dagger$~\cite{alpaca} & 0.000 & 0.065 & 41.989 & 0.004 & 0.024 & 0.008 & 0.138 \\
Baize$^\dagger$~\cite{baize} & 0.000 & 0.044 & 41.500 & 0.004 & 0.025 & 0.009 & 0.097 \\
ChatGLM$^\dagger$~\cite{zeng2023glm-130b} & 0.000 & 0.183 & 40.008 & 0.050 & 0.100 & 0.044 & 0.108 \\
LLama$^\dagger$~\cite{Llama} & 0.000 & 0.020 & 42.002 & 0.001 & 0.002 & 0.001 & 0.039 \\
Vicuna$^\dagger$~\cite{vicuna} & 0.000 & 0.057 & 41.690 & 0.007 & 0.016 & 0.006 & 0.059 \\
Mol-Instruction~\cite{Mol-Instructions} & 0.045 & 0.654 & 27.262 & 0.313 & 0.509 & 0.262 & 1.000 \\
LLama-7b$^*$~\cite{Llama}(LoRA) &0.012	&0.804	&29.947	&0.499	&0.649	&0.407	&1.000 \\
Text+ChemT5~\cite{Text+ChemT5} & 0.454 &0.602 &26.545 &0.729 &0.773 &0.700 &0.851 \\
\textcolor{gray}{MolelcularTransformer} \cite{MolecularTransformer} & 0.0 &0.476 &45.979 &0.761 &0.673 &0.540 &1.000 \\
\midrule
\textbf{InstructMo-G}        & 0.153 &0.906 &20.155	&0.519	&0.717	&0.457	&1.000 \\
\textbf{InstructMol-GS} & \textbf{0.536} &\textbf{0.967}	&\textbf{10.851}	&\textbf{0.776}	&\textbf{0.878}	&\textbf{0.741}	&1.000\\

\midrule[1.1pt]
\rowcolor[RGB]{234, 238, 234}
\multicolumn{8}{l}{\textit{Retrosynthesis}} \\
Alpaca$^\dagger$~\cite{alpaca} & 0.000 & 0.063 & 46.915 & 0.005 & 0.023 & 0.007 & 0.160 \\
Baize$^\dagger$~\cite{baize} & 0.000 & 0.095 & 44.714 & 0.025 & 0.050 & 0.023 & 0.112 \\
ChatGLM$^\dagger$~\cite{zeng2023glm-130b} & 0.000 & 0.117 & 48.365 & 0.056 & 0.075 & 0.043 & 0.046 \\
LLama$^\dagger$~\cite{Llama} & 0.000 & 0.036 & 46.844 & 0.018 & 0.029 & 0.017 & 0.010 \\
Vicuna$^\dagger$~\cite{vicuna} & 0.000 & 0.057 & 46.877 & 0.025 & 0.030 & 0.021 & 0.017 \\
Mol-Instruction~\cite{Mol-Instructions} & 0.009 & 0.705 & 31.227 & 0.283 & 0.487 & 0.230 & 1.000 \\
LLama-7b$^*$~\cite{Llama}(LoRA) &0.000	&0.283	&53.510	&0.136	&0.294	&0.106	&1.000 \\
Text+ChemT5~\cite{Text+ChemT5} & 0.033 &0.314 &88.672 &0.457 &0.469 &0.350 &0.632 \\
\textcolor{gray}{Retroformer-untyped} \cite{Retroformer} & \textbf{0.536} &0.881 &\textbf{10.277} &\textbf{0.865} &\textbf{0.904} &\textbf{0.830} &0.995 \\
\midrule
\textbf{InstructMol-G}       & 0.114	&0.586	&21.271	&0.422	&0.523	&0.285	&1.000\\
\textbf{InstructMol-GS} & 0.407	&\textbf{0.941}	&13.967 & 0.753 &{0.852}	& 0.714	&1.000\\
\bottomrule
\end{tabular}
}}
\caption{
Results of chemical reaction tasks. $\dagger$: few-shot ICL results from \cite{Mol-Instructions}. $*$: use task-specific instruction data to fine-tune. \textcolor{gray}{Model} indicates a domain expert method.
}
\label{tab:chemical_reaction}
\end{table*}

\subsection{Ablation Studies}

In this subsection, we conduct an ablation study to investigate the architecture and training scheme design of our proposed framework. We explore variations from several perspectives and validate them on the task of molecule description generation. The ablation results are presented in Appendix Table~\ref{tab:ablation} as follows: \textbf{1) Employing an MLP connector instead of a linear projector}. Drawing inspiration from the observations made in \cite{llavav1.5}, we attempt to change the alignment projector to a two-layer MLP, demonstrating an enhancement in the model's multimodal capabilities. \textbf{2) Scaling up the LLM to 13B.} The results indicate that scaling up the LLM only yields minor improvements. Thus, it substantiates the assertion that, for specific domains characterized by dataset scarcity, employing a 7B size model is sufficiently efficient for modeling. \textbf{3) Replacing the graph encoder $f_g$ with a single-modality module} (i.e., GraphMVP~\cite{GraphMVP} with the same parameter size and architecture as we used). The results affirm our perspective: utilizing an encoder pre-aligned with text enhances the effectiveness of modality alignment. \textbf{4) Skipping alignment stage-1.} We included a comparison where stage-1 was skipped entirely. The results demonstrate that separating projector training (stage-1) from downstream fine-tuning (stage-2) yields better performance. \textbf{5) Freezing the LLM in the second stage.} Adopting a strategy akin to BioMedGPT10B~\cite{BioMedGPT} and DrugChat~\cite{DrugChat}, we choose not to update LLM weights in the second stage. The training outcomes reveal challenges in convergence and an inability to complete normal inference, thus demonstrating the necessity for the instruct-tuning stage to adapt LLM knowledge to the specific task.

\section{Discussion and Conclusion}
\label{sec:conclusion}
\textbf{Conclusion.} We propose InstructMol, a novel multi-modality foundational model that connects molecular modalities with human natural language. By integrating structural and sequential information of molecules into LLMs through a dual-stage alignment pre-training and instruction tuning paradigm, we enhance the general LLM's capacity to comprehend and interpret molecular information, specifically in drug discovery tasks. Extensive experimental evaluation confirms the effectiveness of our model architecture and training approach, demonstrating its potential for practical applications in the field of drug discovery.

\noindent
\textbf{Future Work.} 
Integrating multiple modalities with LLMs significantly enhances molecular research within this domain and is a valuable direction to explore. However, several challenges exist.
The scale and quality of relevant datasets are as good as those in the vision and language community. The lack of well-defined task objectives poses a challenge. A more scientifically robust evaluation is needed to address issues such as hallucinations in generation outputs.

\section{Limitations}
\label{sec:Limitations} 
In our investigation, several limitations have emerged. Firstly, the scale and quality of the dataset pose significant constraints; the scarcity of high-quality annotated domain data may hinder the model's ability to generalize across the diverse and intricate molecular landscapes encountered in real-world applications. Secondly, the integration and evaluation of multiple modalities have also revealed areas needing improvement. Further refinement is necessary to ensure robust alignment and utilization of different molecule modalities within the model, enhancing its capacity to interpret and generate responses accurately across the molecular domain. Lastly, our base LLM originates from a general-domain model. However, the absence of specialized LLMs tailored specifically for chemistry and molecular science, like models such as LLaMA, highlights the need for larger, more versatile domain-specific LLMs to enhance performance and expand applications. Addressing these challenges is pivotal for enhancing the model's reliability and extending its utility in advancing drug discovery methodologies.

\section{Potential Risks}
\label{sec:Ethical}
The application of AI in drug discovery entails several potential risks. A primary concern is the potential misuse of AI to develop hazardous or illicit substances, which presents significant safety and ethical challenges. Moreover, inaccuracies in AI-generated outputs could lead to hazardous chemical reactions if not thoroughly verified, posing risks of harm or damage to equipment. Dependence on AI-generated content heightens the risk of accidents and unsafe practices. Therefore, stringent oversight and rigorous adherence to ethical guidelines are essential to mitigate these risks and ensure the safe and responsible application of AI in drug discovery.
Further insights into these issues and potential safeguard approaches can be found in recent literature~\cite{wong2024smiles,cao2024guide,wang2024adashield}.

\section{Acknowledgements}
\label{sec:ack}
This project was supported in part by Shenzhen Hetao Shenzhen-Hong Kong Science and Technology Innovation Cooperation Zone, under Grant No. HTHZQSWS-KCCYB-2023052, National Natural Science Foundation of China / Research Grants Council Joint Research Scheme Grant N\_HKUST635/20, and HKRGC Grant 16308321.

\bibliography{main}

\clearpage
\appendix
\label{appendix}

\section{Tasks Definition and Dataset Details}
\label{appendix:task-def}

\begin{table*}[ht]
\small
\vskip -0.07in
\centering
\setlength{\tabcolsep}{6mm}{
\scalebox{0.92}{
\begin{tabular}{lcc}
    \toprule
    \textsc{tasks} & \textsc{\# Samples} & \textsc{Data Source} \\
    \midrule
    Alignment Pretrain & 264K & PubMed~\cite{PubChem} \\
    Property Prediction(Regression) & 362K & QM9~\cite{Mol-Instructions, MoleculeNet} \\ 
    Property Prediction(Classification) &35,742 & BACE, BBBP, HIV~\cite{MoleculeNet} \\
    Molecule Description Generation & 26,507 & ChEBI-20~\cite{Text2Mol} \\
    Forward Prediction & 125K & USPTO~\cite{Mol-Instructions, USPTO} \\ 
    Retrosynthesis & 130K & USPTO\_500MT~\cite{Mol-Instructions, USPTO_500MT} \\ 
    Reagent Prediction & 125K & USPTO\_500K~\cite{Mol-Instructions, USPTO_500MT} \\
    \bottomrule
    
\end{tabular}
}}
\caption{\footnotesize
Details of InstrutMol two-stage training data.
}
\vskip -0.1in
\label{tab:data_collection}
\end{table*}

\paragraph{Property Prediction.}
Molecular Property Prediction involves the forecasting or estimation of the biophysical and chemical properties of a molecule. 
In this work, our emphasis lies on three binary classification tasks sourced from the MoleculeNet benchmark (BBBP, BACE, and HIV)~\cite{MoleculeNet}, and three regression tasks concentrating on the quantum properties of molecules from the QM9~\cite{QM9} dataset.

\paragraph{Molecule Description Generation.}
Generating molecular descriptions involves compiling a detailed overview of a molecule's structure, properties, activities, and functions. This process aids chemists and biologists by swiftly providing crucial molecular insights for their research. Our data collection involves the extraction of molecular text annotations from PubChem~\cite{PubChem}. Leveraging PubChem's \textbf{Power User Gateway}~\cite{PUG-View}, we retrieve abstracts of compound records in XML format. Subsequently, we extracted valid molecular description texts identified by unique PubChem Chemical Identifiers (CIDs), filtering out SMILES strings with syntactic errors or deviations from established chemical principles. Furthermore, we utilize the ChEBI-20 dataset~\cite{Text2Mol} for downstream tasks in molecule description generation, comprising 33,010 molecule description pairs divided into 80\% for training, 10\% for validation and 10\% for testing. To prevent data leakage, compounds in the PubChem text annotations that coincide with the ChEBI-20 test split are excluded.

\paragraph{Forward Reaction Prediction.}
Predicting the forward reaction involves anticipating the probable product(s) of a chemical reaction based on given reactants and reagents. 
For this task, we utilize the forward-reaction-prediction dataset from \cite{Mol-Instructions}, comprising 138,768 samples sourced from the USPTO dataset~\cite{USPTO}. Each entry includes reactants and reagents separated by '.' within the instruction, with the output product.

\paragraph{Reagent Prediction.}
Reagent prediction identifies the substances necessary for a chemical reaction, helping to discover new types of reaction and optimal conditions. We use the reagent Prediction data from~\cite{Mol-Instructions}, sourced from the USPTO\_500MT dataset~\cite{USPTO_500MT}. Each entry features a chemical reaction indicated as ``reactants $>>$ product," with the output indicating the reagents involved in the reaction.

\paragraph{Retrosynthesis Prediction.}
Retrosynthetic analysis in organic chemistry reverses engineering by tracing potential synthesis routes from the target compound backward. This strategy is vital for efficient synthesis of complex molecules and to foster innovation in pharmaceuticals and materials. For this task, we also used the dataset from \cite{Mol-Instructions}, which is sourced from USPTO\_500MT. The data organize inputs as products and outputs as reactants separated by '.' for each compound.

\paragraph{Discussion on License.} 
As depicted in Table~\ref{tab:license}, we elaborate on the origins and legal permissions associated with each data component utilized in the development of the InstructMol. This encompasses both biomolecular data and textual descriptions. Thorough scrutiny was conducted on all data origins to confirm compatibility with our research objectives and subsequent utilization. Proper and accurate citation of these data sources is consistently maintained throughout the paper.

\begin{table*}[ht]
    \centering
    \vskip 0.1in
    \scalebox{0.69}{
    \begin{tabular}{p{4cm}p{7cm}p{11cm}}
    \toprule
    \textsc{\textbf{Data Sources}} & \textsc{\textbf{License URL}} & \textsc{\textbf{License Note}} \\
    \midrule
    PubChem & \url{https://www.nlm.nih.gov/web_policies.html} & Works produced by the U.S. government are not subject to copyright protection in the United States. Any such works found on National Library of Medicine (NLM) Web sites may be freely used or reproduced without permission in the U.S. \\
    ChEBI & \url{https://creativecommons.org/licenses/by/4.0/} & You are free to: Share — copy and redistribute the material in any medium or format. Adapt — remix, transform, and build upon the material for any purpose, even commercially. \\
    USPTO & \url{https://www.uspto.gov/learning-and-resources/open-data-and-mobility} & It can be freely used, reused, and redistributed by anyone. \\
    MoleculeNet & \url{https://opensource.org/ license/mit/} & Permission is hereby granted, free of charge, to any person obtaining a copy of this software and associated documentation files (the “Software”), to deal in the Software without restriction, including without limitation the rights to use, copy, modify, merge, publish, distribute, sublicense, and/or sell copies of the Software, and to permit persons to whom the Software is furnished to do so.\\
    \bottomrule
    \end{tabular}
    }
    \caption{
    \footnotesize{
    Data resources and licenses utilized in data collection.}.
    }
    \vskip -0.1in
    \label{tab:license}
\end{table*}

\section{Implementation Details}
\label{sec:implementation_details}

\textbf{Model Settings.}
A graph neural network with five graph isomorphism network (GIN)~\cite{GIN} layers is used as the molecule graph encoder $f_g$. The hidden dimension is set to be 300. The GIN model is initialized using the MoleculeSTM~\cite{moleculeSTM} graph encoder, which is pre-trained through molecular graph-text contrastive learning. We employ Vicuna-v-1.3-7B~\cite{vicuna} as the base LLM, which has been trained through instruction-tuning. The total number of parameters of InstructMol is around 6.9B. 

\noindent
\textbf{Training Details.} In the first stage, we employ the training split comprising around 264K molecule-caption pairs from PubMed. Using a batch size of 128, we conduct training for 5 epochs. We use the AdamW optimizer, with $\beta$=(0.9, 0.999) and a learning rate of 2e-3, without weight decay. Warm-up is executed over 3\% of the total training steps, followed by a cosine schedule for learning rate decay. 
For the second stage, we conduct training for three specific scenarios. For fair comparisons with traditional methods, training spans 20 to 50 epochs for the molecule description generation task using the ChEBI-20 training split. Property prediction and reaction tasks undergo 10 epochs using corresponding instruction datasets. In InstructMol training, we maintain a consistent batch size of 128 and set the learning rate to 8e-5. Linear layers within the LLM utilize a LoRA rank of 64 and a scaling value $\alpha$ of 16. All experiments are run with 4$\times$RTX A6000 (48GB) GPUs.

\begin{table}[h]
\small
\centering
\renewcommand\arraystretch{1.0}
\setlength{\tabcolsep}{5mm}
\scalebox{1.0}{
\begin{tabular}{lc}
\toprule
\textbf{Configuration} & \textbf{Value} \\
\midrule
    Graph encoder $f_g$ init. & $\text{GIN}_{\text{MoleculeSTM}}$ \\
    \# params $f_g$ & 1.8M \\
    LLM init. & Vicuna-v-1.3-7B \\
    \# params LLM & 6.9B \\
    Stage1 batch-size & 128 \\
    Stage2 batch-size & 128 \\
    Optimizer & AdamW \\
    Warm-up ratios & 0.03 \\
    Stage1 peak lr & 2e-3 \\
    Stage2 peak lr & 8e-5 \\
    Learning rate schedule & cosine decay \\
    Weight decay & 0. \\
    Stage1 train epochs & 5 \\
    Stage2 train epochs & 20-50 \\
    Numerical precision & bfloat16 \\
    Activation checkpointing & True \\
\bottomrule
\end{tabular}
}
\caption{\footnotesize{Training hyperparameters of InstructMol.}}
\label{training_detail}
\end{table}

\section{Evaluate Metrics}
\label{appendix:eval-metric}
\paragraph{Molecule Description Generation Metric.} 
Following \cite{MolT5}, NLP metrics such as BLEU~\cite{BLEU}, ROUGE~\cite{ROUGE} and METEOR~\cite{METEOR} are used to assess the proximity of generated descriptions to the truth of the ground. Specifically, these metrics are tested on the ChEBI-20 test dataset. In our experiments, we observed that after 50 epochs of finetuning on the training split, the metrics tend to converge, differing from previous approaches that often involved fine-tuning for over 100 epochs~\cite{MolT5, MoMu, MolFM}.

\paragraph{Molecule Generation Metric.}
In chemical reaction tasks, we view it as akin to a text-based molecule generation task. Initially, we employ RDKit to validate the chemical validity of the generated results, ensuring their ``validity". Subsequently, we gauge the sequential proximity between the generated sequence and the ground truth using NLP metrics such as BLEU, Exact Match scores, and Levenshtein distance. Additionally, we present performance based on molecule-specific metrics that assess molecular similarity, encompassing RDKit, MACCS~\cite{MACCS}, and Morgan~\cite{Morgan} fingerprints similarity. 

\begin{table*}[ht]
\centering
\footnotesize
\setlength{\tabcolsep}{2.0mm}{
\scalebox{1.0}{
\begin{tabular}{ll}
\toprule
\textbf{\textsc{Task}} & \textbf{\textsc{Instruction}} \\ \hline
Alignment Pretrain  & \makecell[l]{Instruction: \textit{Provide a brief overview of this molecule.}\\ \textit{$\mathbin\Vert\ [$Optional: The compound SELFIES sequence is: } \texttt{SELFIES}$]$ \\ Output: \textit{The molecule is a non-proteinogenic alpha-amino acid that is ...} } \\ 
\hline
\makecell[l]{Property Prediction \\(Regression)}  &  \makecell[l]{Instruction: \textit{Could you give me the LUMO energy value of this molecule?}\\ \textit{$\mathbin\Vert\ [$Optional: The compound SELFIES sequence is: } \texttt{SELFIES}$]$ \\ Output: \textit{0.0576} } \\
\hline
\makecell[l]{Property Prediction \\(Classification)} & \makecell[l]{Instruction: \textit{Evaluate whether the given molecule is able to enter the blood-brain barrier.}\\ \textit{$\mathbin\Vert\ [$Optional: The compound SELFIES sequence is: } \texttt{SELFIES}$]$ \\ Output: \textit{Yes} } \\
\hline
Molecule Description Generation & \makecell[l]{Instruction: \textit{Could you give me a brief overview of this molecule?}\\ \textit{$\mathbin\Vert\ [$Optional: The compound SELFIES sequence is: } \texttt{SELFIES}$]$ \\ Output:\textit{The molecule is a fatty acid ester obtained by ...}  } \\
\hline
Forward Prediction & \makecell[l]{Instruction: \textit{Based on the given reactants and reagents, suggest a possible product.}\\ $\mathbin\Vert\ $\texttt{<REACTANT A>.<REACTANT B>...<REAGENT A>.<REAGENT B>...} \\ Output: \texttt{SELFIES} of product } \\
\hline
Retrosynthesis &  \makecell[l]{Instruction: \textit{Please suggest potential reactants used in the synthesis of the provided product.}\\ $\mathbin\Vert\ $\texttt{SELFIES} of product \\ Output: \texttt{<REACTANT A>.<REACTANT B>...<REAGENT A>.<REAGENT B>...}  } \\
\hline
Reagent Prediction  &  \makecell[l]{Instruction: \textit{Can you provide potential reagents for the following chemical reaction?}\\ $\mathbin\Vert\ $ \texttt{<REACTANT A>.<REACTANT B>...<REAGENT A>.<REAGENT B>... >> <PRODUCTs>}  \\ Output: \texttt{SELFIES} of reagent }  \\
\bottomrule
\end{tabular}
}}
\caption{
    Examples of instruction samples for each task. $\mathbin\Vert$ means concatenate along the token dimension.
}
\label{tab:appendix-dataset}
\end{table*}

\vskip 2in
\begin{table*}[h!]\centering
\begin{minipage}{\textwidth}\vspace{0mm}    
\centering
\begin{tcolorbox} 
    \centering
    \small
\begin{tabular}{p{0.99\textwidth}}

\VarSty{messages} = [
            \{\var{"role":"system", "content":} \var{f"""}You're acting as a molecule property prediction assistant. You'll be given SMILES of molecules and you need to make binary classification with a return result only in ``True" or ``False". \\
            \\ 
\textbf{The background of the dataset and task is shown below:} \\
The Blood-brain barrier penetration (BBBP) dataset comes from a recent study on the modeling and prediction of barrier permeability. As a membrane separating circulating blood and brain extracellular fluid, the blood-brain barrier blocks most drugs, hormones, and neurotransmitters. Thus penetration of the barrier forms a long-standing issue in the development of drugs targeting the central nervous system.   \\
\\
We provide several examples for this binary classification task: \\
\#\#\# \\
Instruction: Predict whether the given compound has barrier permeability. Return True or False.\\ 
SMILES: CCC(=O)C(CC(C)N(C)C)(c1ccccc1)c2ccccc2 \\
Output: True \\
\#\#\# \\
\\
\#\#\# \\
Instruction: Predict whether the provided compound exhibits barrier permeability. Return True or False.\\ 
SMILES: c1cc2c(cc(CC3=CNC(=NC3=O)NCCSCc3oc(cc3)CN(C)C)cc2)cc1 \\
Output: False \\
\#\#\# \\
... \\
\\

Given the following instructions and SMILES, return your prediction result: \\
Instruction: Predict whether the provided compound exhibits barrier permeability. Return True or False. \\
SMILES: \var{TARGET SMILES} \\

\var{"""}\}\\
]
\\
\end{tabular}
\end{tcolorbox}
\vspace{-2mm}
\caption{An illustration of the few-shot in-context-learning prompt construction process for Llama~\cite{Llama, Llama2} / Vicuna~\cite{vicuna} models in property prediction tasks.}
\label{tab:appendix-ICL-prompt}
\end{minipage}
\end{table*}

\clearpage
\onecolumn
\section{More Results}
\label{appendix:more-examples}

\subsection{Ablation study results}
\begin{table*}[!htp]
\centering
\small
\setlength{\tabcolsep}{0.6mm}{
\scalebox{0.95}{
\begin{tabular}{lcccccc}
\toprule
\textsc{Methods}
&\textsc{BLEU-2}$\uparrow$  & \textsc{BLEU-4}$\uparrow$  & \textsc{ROUGE-1}$\uparrow$  & \textsc{ROUGE-2}$\uparrow$  & \textsc{ROUGE-L}$\uparrow$ & \textsc{METEOR}$\uparrow$ \\
\midrule[1.1pt]
\textbf{InstructMol-G}         &0.4620	&0.3560	&0.5439	&0.3644	&0.4765	&0.4832	\\
+MLP XL connector   &\textbf{0.4665}(\footnotesize{\textcolor{mygreen}{+0.97\%}}) &\textbf{0.3613}(\footnotesize{\textcolor{mygreen}{+1.49\%}}) &\textbf{0.5497}(\footnotesize{\textcolor{mygreen}{+1.07\%}}) &\textbf{0.3699}(\footnotesize{\textcolor{mygreen}{+1.51\%}}) &\textbf{0.4805}(\footnotesize{\textcolor{mygreen}{+0.84\%}}) &\textbf{0.4917}(\footnotesize{\textcolor{mygreen}{+1.76\%}}) \\
+Scale up LLM       &0.4615(\footnotesize{\textcolor{red}{-0.11\%}}) &0.3566(\footnotesize{\textcolor{mygreen}{+0.17\%}}) &0.5449(\footnotesize{\textcolor{mygreen}{+0.18\%}}) &0.3660(\footnotesize{\textcolor{mygreen}{+0.44\%}}) &0.4776(\footnotesize{\textcolor{mygreen}{+0.23\%}}) &0.4868(\footnotesize{\textcolor{mygreen}{+0.75\%}}) \\
Replace $f_g$ with GraphMVP & 0.4452(\footnotesize{\textcolor{red}{-3.64\%}}) & 0.3377(\footnotesize{\textcolor{red}{-5.14\%}}) & 0.5318(\footnotesize{\textcolor{red}{-0.11\%}}) & 0.3484(\footnotesize{\textcolor{red}{-2.22\%}}) & 0.4638(\footnotesize{\textcolor{red}{-2.67\%}}) & 0.4691(\footnotesize{\textcolor{red}{-2.92\%}}) \\
Skip Stage-1 & 0.4631(\footnotesize{\textcolor{mygreen}{+0.23\%}}) & 0.3569(\footnotesize{\textcolor{mygreen}{+0.25\%}}) & 0.5419(\footnotesize{\textcolor{red}{-0.37\%}}) & 0.3610(\footnotesize{\textcolor{red}{-0.93\%}}) & 	0.4720(\footnotesize{\textcolor{red}{-0.94\%}}) & 0.4391(\footnotesize{\textcolor{red}{-9.12\%}}) \\
Freeze LLM in the second stage & $\sim 0$ & $\sim 0$ & $\sim 0$ & $\sim 0$ & $\sim 0$ & $\sim 0$ \\
\bottomrule
\end{tabular}
}}
\caption{\small
Ablation of the model architecture and training scheme design. We chose to conduct experiments on the Molecule Description Generation task. $f_g$ represents the molecule graph encoder. 
}
\label{tab:ablation}
\end{table*}

\subsection{More Results of Molecule Description Generation}
\begin{figure*}[!ht]
    \centering
    \includegraphics[width=1\linewidth]{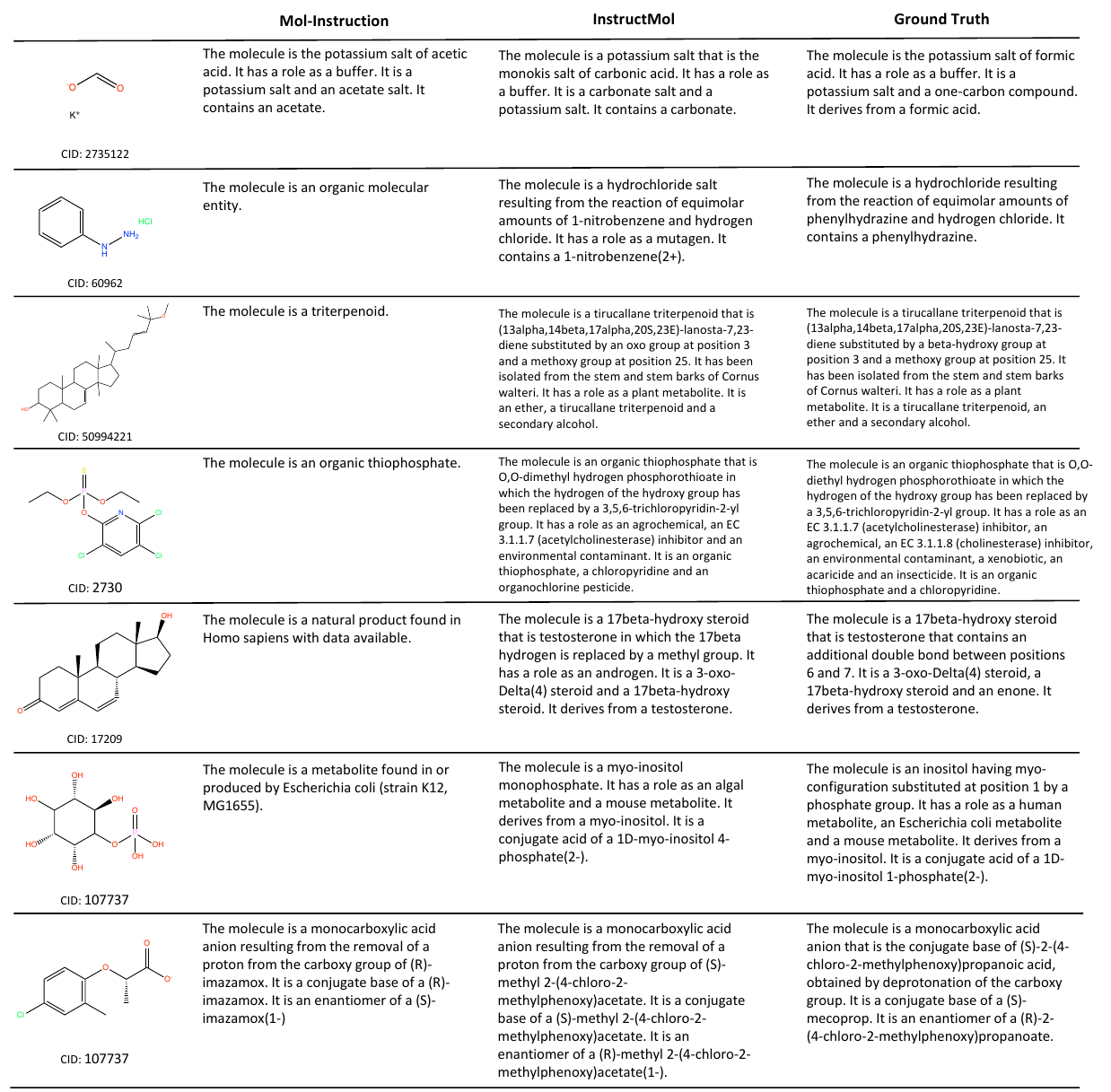}
    \caption{More examples of molecule description generation task on ChEBI-20~\cite{Text2Mol} test set. We include Mol-Instruction~\cite{Mol-Instructions} as the baseline. CID~\cite{CID}: PubChem Compound Identification, a non-zero integer PubChem accession identifier for a unique chemical structure. }
    \label{fig:appendix-molcap}
\end{figure*}

\clearpage
\subsection{More Results of Forward Reaction Prediction}
\begin{figure*}[!ht]
    \centering
    \includegraphics[width=1\linewidth]{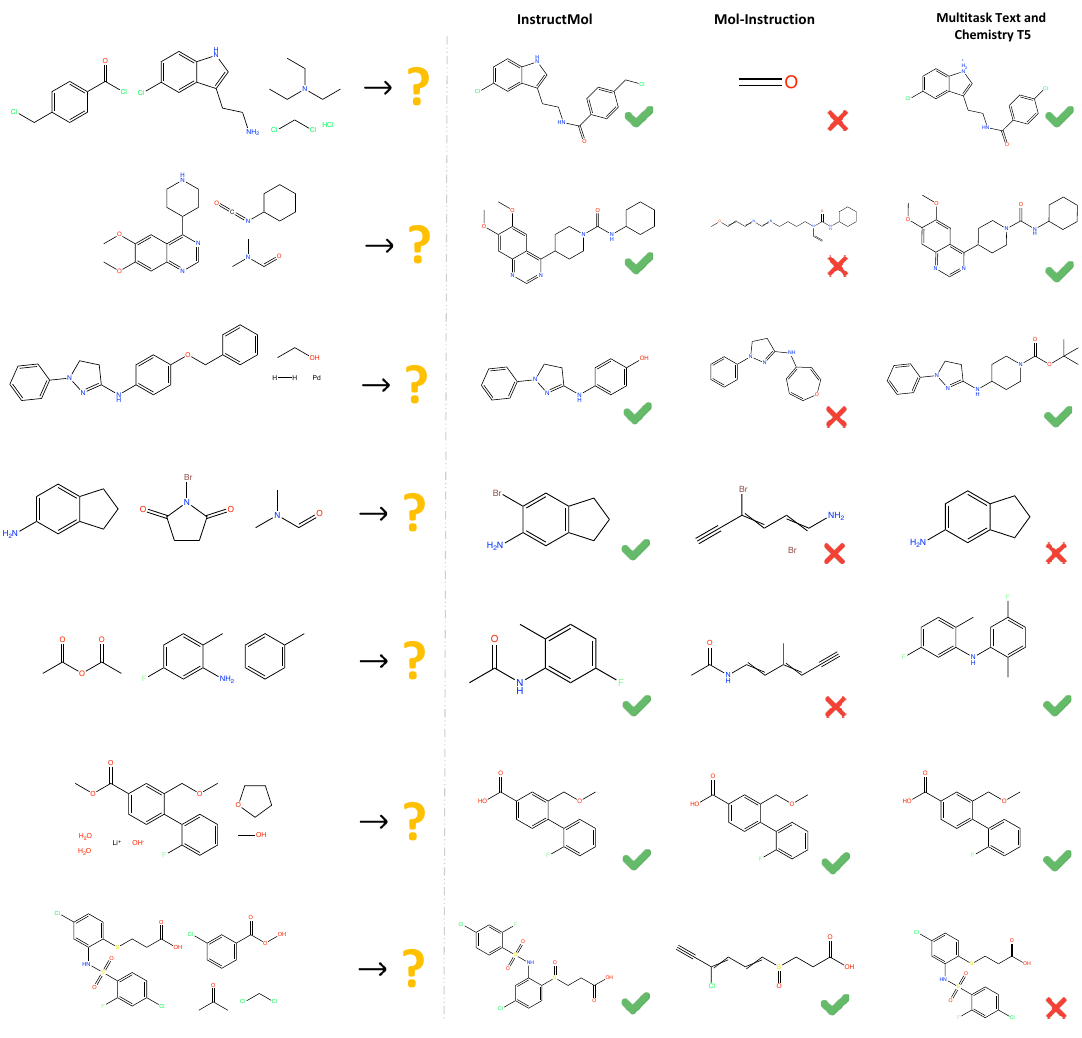}
    \caption{More examples of forward reaction prediction task. We include Mol-Instruction~\cite{Mol-Instructions} and Multitask-Text-and-Chemistry-T5~\cite{Text+ChemT5} as baselines.}
    \label{fig:appendix-forward}
\end{figure*}

\clearpage
\subsection{More Results of Reagent Prediction}
\begin{figure*}[!ht]
    \centering
    \includegraphics[width=1\linewidth]{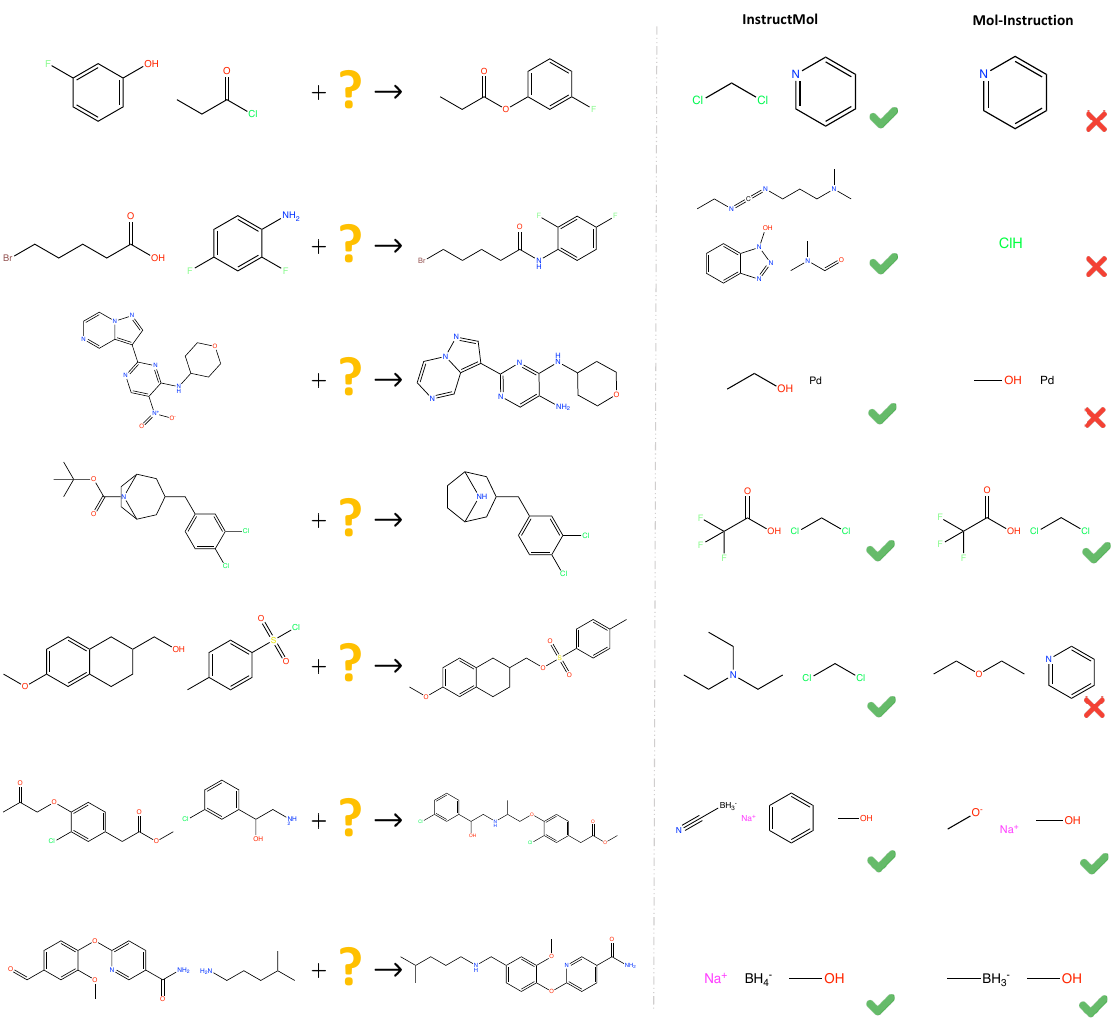}
    \caption{More examples of the reagent prediction task. We include Mol-Instruction~\cite{Mol-Instructions} as the baseline.}
    \label{fig:appendix-reagent}
\end{figure*}

\clearpage
\subsection{More Results of Retrosynthesis Prediction}
\begin{figure*}[!ht]
    \centering
    \includegraphics[width=1\linewidth]{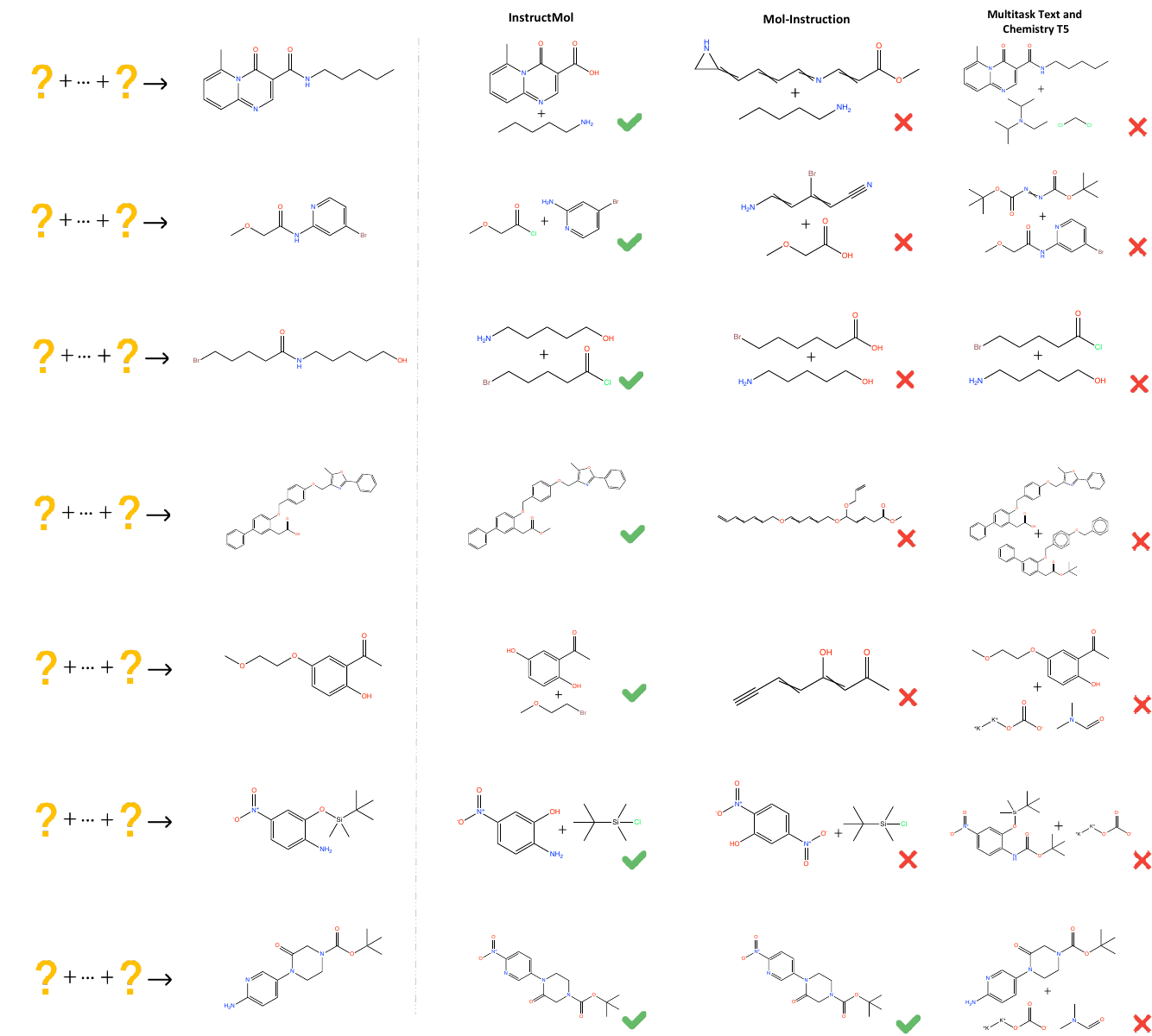}
    \caption{More examples of the retrosynthesis prediction task. We include Mol-Instruction~\cite{Mol-Instructions} and Multitask-Text-and-Chemistry-T5~\cite{Text+ChemT5} as baselines.}
    \label{fig:appendix-retro}
\end{figure*}

\clearpage
\twocolumn
\subsection{Error Analysis}
\begin{figure*}[!ht]
    \centering
    \includegraphics[width=1\linewidth]{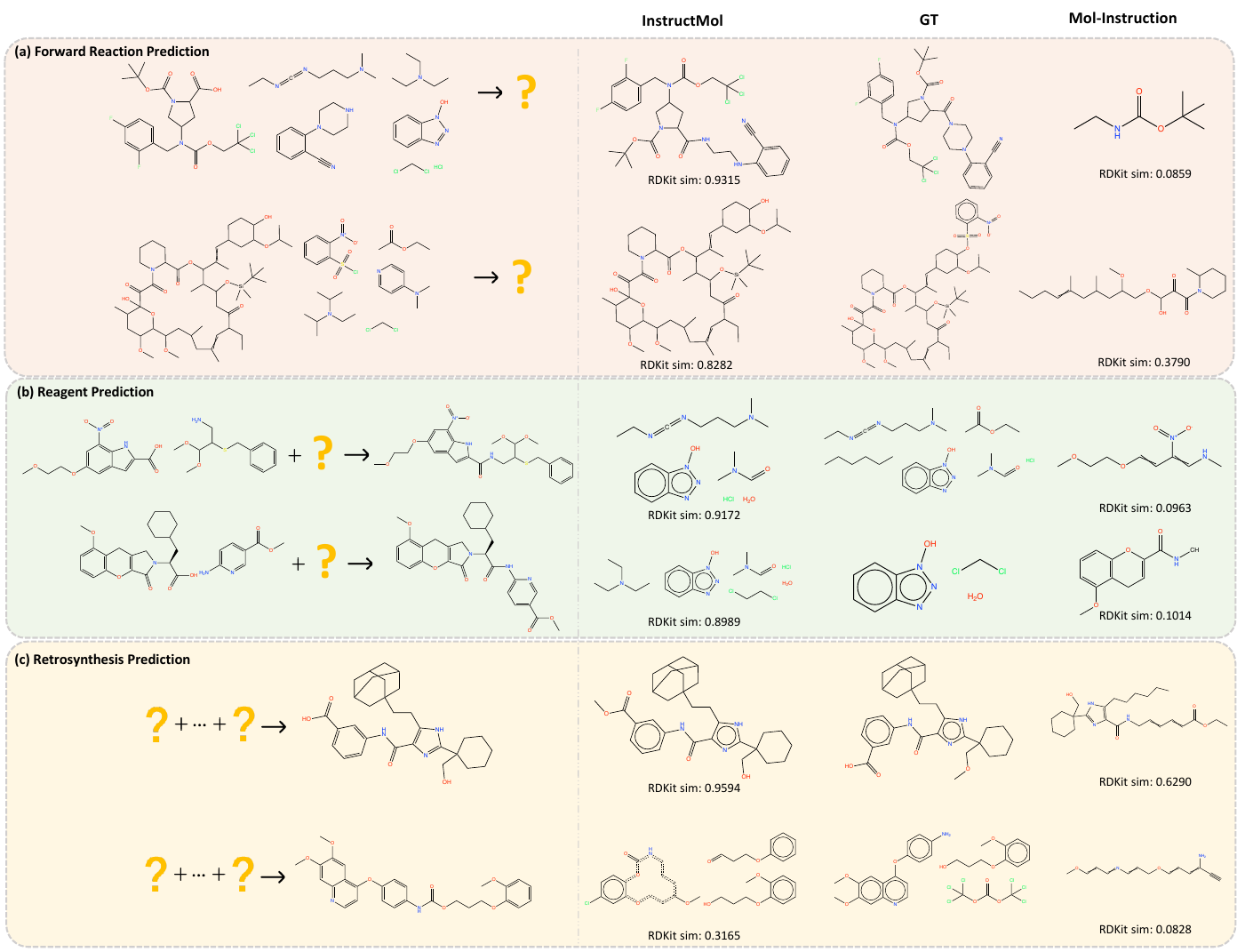}
    \caption{We present several cases with a certain degree of misalignment compared to the ground truth, accompanied by RDKit fingerprint similarity results relative to the ground truth. Due to the heightened complexity of compounds involved in chemical reactions, the difficulty of the task increases, leading to the poor performance of Mol-Instructions~\cite{Mol-Instructions}.}
    \label{fig:appendix-failure}
\end{figure*}

We showcase cases with misalignment to the ground truth, along with RDKit fingerprint similarity results in Fig.~\ref{fig:appendix-failure}. The complexity of chemical reaction compounds makes the task more challenging. In addressing this limitation, our future approach involves concatenating graph tokens from multiple molecules involved in the same reaction with text tokens to simplify the complexity of the input sequence. Moreover, we are considering employing separate tokenization and embedding for distinct modalities to ensure the semantic accuracy of the tokenized results.

\subsection{More Results of Molecule Property Prediction}
Based on the molecular property binary classification discussed in the main text, we have extended our analysis by comparing Instruct-G with other domain-specific models (ChemBERTa~\cite{ChemBERTa}, GROVER~\cite{GROVER}, DMP~\cite{DMP}) and MolCA variants on multi-class prediction tasks from MoleculeNet, including Clintox, Tox21, Toxcast, and SIDER, as detailed in the Table~\ref{tab:multi-classification}.

\begin{table}[!ht]
    \centering
    \resizebox{1.05\linewidth}{!}{
    \begin{tabular}{lcccc}
        \toprule
        \textsc{Method} &Clintox  &Tox21 &Toxcast & SIDER \\
        \textsc{\# Molecules} &1491 &8014 &8615 &1427 \\
        \textsc{\# Tasks} & 2 & 12 & 617 & 27 \\
        \midrule
        \rowcolor[RGB]{234, 238, 234} \multicolumn{5}{l}{\textit{Specialist Models}} \\
        ChemBERTa~\cite{ChemBERTa} & 	73.3	&-	&-	&- \\
        ChemBERTa2~\cite{chemberta2} & 23.9	&-	&-	&- \\
        GROVER-large~\cite{GROVER} &94.4	&83.1	&73.7	&65.8 \\
        DMP(TF+GNN)~\cite{DMP} &95.6	&79.1	&-	&69.8 \\
        MolCA(1D+2D)~\cite{MolCA} &89.5	&77.2	&64.5 & - \\
        \midrule 
        \textbf{Instruct-G} & 93.4 & 77.0 & 61.9 & 62.2 \\
        \bottomrule
    \end{tabular}
    }
    \caption{\small{
    ROC-AUC results of molecular property tasks (multi-classes classification) on MoleculeNet~\cite{MoleculeNet} benchmarks.}}
    \label{tab:multi-classification}
\end{table}

Based on the results of Table \ref{tab:property_pred_classification} and Table \ref{tab:multi-classification}, we found that integrating the 1D and 2D molecular modalities significantly enhances the model's understanding capabilities. It is important to note that ChemBERTa, GROVER, and DMP were all pre-trained on large molecule-only datasets: ChemBERTa on 77M unique SMILES, GROVER on 11M molecules, and DMP on 110M molecules. In contrast, InstructMol utilized only about 300K molecule-text description pairs for the initial alignment stage, with parameter size updates confined to the projector layer ($<1$ million), and without extensive retraining. This limited the molecule space it covered. To further improve performance on MoleculeNet, additional pretraining stages and the collection of large unlabeled datasets to cover a broader range of molecules could be considered.

\subsection{From LoRA to Full-Finetuning}
InstructMol is instruction-tuned using LoRA, with a trainable parameter size of less than 100M, which is significantly lower than that of domain expert models like the MolT5~\cite{MolT5} series. These domain expert models are pretrained on over 100 million SMILES and are limited to only a few tasks, such as molecule captioning and de novo design. The main focus of our work is to demonstrate that the aligning SFT training approach can efficiently and rapidly adapt general language models into domain-specific multimodal models capable of addressing multiple downstream tasks. Increasing the trainable parameters and adding additional pretraining datasets will further boost InstructMol's performance, as shown in Table~\ref{tab:full-tune}.

\begin{table*}[!htp]
    \centering
    \small
    \setlength{\tabcolsep}{0.6mm}{
    \scalebox{0.95}{
    \begin{tabular}{lcccccc}
    \toprule
    \textsc{Methods}
    &\textsc{BLEU-2}$\uparrow$  & \textsc{BLEU-4}$\uparrow$  & \textsc{ROUGE-1}$\uparrow$  & \textsc{ROUGE-2}$\uparrow$  & \textsc{ROUGE-L}$\uparrow$ & \textsc{METEOR}$\uparrow$ \\
    \midrule
    MolT5-large~\cite{MolT5} &0.594	&0.508	&0.654	&0.510	&0.594	&0.614 \\
    Text+Chem T5 (augm-base)~\cite{Text+ChemT5}	&0.625	&0.542	&0.682	&0.543	&0.622	&0.648 \\
    MolCA~\cite{MolCA} &0.620 &0.531 &0.681 &0.537 & 0.618 & 0.651 \\
    \textbf{Instruct-G} (Full-tune)	&0.653	&0.566	&0.608	&0.445	&0.541	&0.562 \\
    \bottomrule
    \end{tabular}
    }}
    \caption{\small
    Comparison with state-of-the-art models on the Molecule Caption task when performing full fine-tuning.
    }
    \label{tab:full-tune}
\end{table*}

To assess whether InstructMol can retain the original capabilities of LLMs, we conducted additional dialogues using InstructMol. Our findings indicate that the model continues to exhibit communication skills, common sense inference, and logical reasoning at a qualitative level. Additionally, we provided quantitative results on several MMLU tasks~\cite{MMLU} (zero-shot) in Table~\ref{tab:model_performance_combined}, demonstrating that despite the inevitable forgetting problem introduced by fine-tuning, InstructMol retains most of the original LLM's capabilities. 

\begin{table}[h]
    \renewcommand\arraystretch{1.5} 
    \centering
    \resizebox{1.05\linewidth}{!}{
    \begin{tabular}{lccc|ccc}
        \hline
        \textbf{Model} & \multicolumn{3}{c|}{\textbf{High School}} & \multicolumn{3}{c}{\textbf{College}} \\
        & \textbf{Biology} & \textbf{Physics} & \textbf{Chemistry} & \textbf{Biology} & \textbf{Physics} & \textbf{Chemistry} \\
        \hline
        Vicuna-7B-v1.3 & 0.529 & 0.291 & 0.345 & 0.465 & 0.186 & 0.270 \\
        \hline
        InstructMol-GS & 0.481 & 0.258 & 0.246 & 0.438 & 0.196 & 0.230 \\
        \hline
    \end{tabular}
    }
    \caption{Performance comparison of LoRA-tuned models with original models across MMLU high school and college science subjects.}
    \label{tab:model_performance_combined}
\end{table}

\onecolumn
\clearpage
\section{Comparison with Current Agents Framework}
\label{appendix:Comparison_with_agents}
LLMs face a major limitation in performing basic mathematical and chemical operations, which makes handling hallucinations challenging. However, their self-supervised pre-training on diverse knowledge equips them with a strong understanding and reasoning abilities that can be directly applied to new domains. Presenting LLMs as automated assistants offers a programming-free interface for non-experts to leverage their existing capabilities. Agent/assistant paradigms enable the optimal utilization of LLMs' knowledge without the need for specialized model development. For instance, ChemCrow \cite{ChemCrow} is an agent system based on GPT-4 that integrates various chemical tools for solving diverse tasks. We conducted a comparison of three downstream tasks between InstructMol and ChemCrow, and the results are presented in Table \ref{tab:LLMAgents}.

During testing, we observed that ChemCrow's performance is heavily reliant on prompt construction, resulting in unstable output results. For instance, in retrosynthesis planning experiments, we found that agents often misidentify the user's query product as controlled chemistry and refuse to provide an answer. Similarly, in the property prediction task, GPT-4 itself lacks specific knowledge about compounds and thus heavily relies on internet searches. The quality of the prompt constructed by the user significantly influences the quality of the response.

\begin{table*}[h]
\centering
\renewcommand\arraystretch{1.0} 
\resizebox{1.0\linewidth}{!}{
\begin{tabular}{lccc}
\hline
\textbf{Task} & \textbf{Ground Truth} & \textbf{ChemCrow} & \textbf{InstructMol}\\
\hline
\rowcolor[RGB]{234, 238, 234}
\multicolumn{4}{l}{\textit{Property Prediction}} \\
\makecell[l]{Determine whether (CID:219214)\\ can suppress HIV.} & "Active" & \makecell[l]{WebSearch$\rightarrow$ \\ No information} & $\checkmark$ \\ \hline
\rowcolor[RGB]{234, 238, 234}
\multicolumn{4}{l}{\textit{Forward Reaction Prediction}} \\
\scriptsize{CCC(=O)Cl $+$ OC1=CC=CC(F)=C1}\\ \scriptsize{$+$ ClCCl $+$C2=CC=NC=C2 $\rightarrow ?$}  & \scriptsize{CCC(=O)OC1=CC=CC(F)=C1} & $\checkmark$  & $\checkmark$  \\ \hline

\rowcolor[RGB]{234, 238, 234}
\multicolumn{4}{l}{\textit{Retrosynthesis Prediction}} \\ 
$?\rightarrow$ \scriptsize{C(CCNC(=O)CCCCBr)CCO} & \scriptsize{NCCCCCO.O=C(O)CCCCCBr} & \makecell[l]{"Similar to controlled \\chemistry, reject to answer"} & $\checkmark$  \\
\bottomrule
\end{tabular}
}
\caption{\footnotesize{The performance of InstructMol and ChemCrow was evaluated through a comparison of three downstream tasks: Property Prediction, Forward Reaction Prediction, and Retrosynthesis. The $\checkmark$ denotes that the predictions match with the ground truths.}}
\label{tab:LLMAgents}
\end{table*}

Therefore, we believe that domain-specific LLMs should be augmented with dedicated external tools. This augmentation would enable LLMs to function as planners, comprehend and decompose tasks, invoke downstream interfaces, and effectively process feedback. In our future work, we intend to create a new dataset for instruction-following tool usage and enhance InstructMol with a variety of external tools. By leveraging state-of-the-art models and maximizing LLM's reasoning and planning capabilities, we aim to further enhance its performance.

\end{document}